\documentclass[aps,prl,twocolumn,superscriptaddress]{revtex4}
 \RequirePackage{lineno}
\usepackage{amsmath,bm}
\usepackage{mathrsfs}
\usepackage{amsfonts}
\usepackage{graphicx}
\usepackage{setspace} %leaves captions single space in draft mode

\usepackage{epstopdf}
\usepackage{dcolumn}

\usepackage{epsfig}
\usepackage{indentfirst}
\usepackage{psfrag}
\usepackage{subfigure}
\usepackage{amssymb}
\usepackage{color}

\usepackage{dcolumn}% Align table columns on decimal point
\usepackage{bm}% bold math

\usepackage{physics}
\usepackage{bm}% bold math
\usepackage{hyperref}% add hypertext capabilities

 \usepackage{graphicx}% Include figure files
 \usepackage{dcolumn}% Align table columns on decimal point
 \usepackage{bm}% bold math

 \def\didv{\mbox{$d\mbox{I}/d\mbox{V$_{\textrm{s}}$}$}}
 \def\vs{\mbox{V$_{\textrm{s}}$}}
 
 \def\It{\mbox{I$_{\textrm{t}}$}}

 \def\vs{V$_s$}
 \def\It{I$_t$}


\begin{document}
 \title{On-surface route for producing planar nanographenes with  azulene moieties}
 
 \author{Jeremy Hieulle}
   \affiliation{CIC nanoGUNE,  20018 San Sebasti\'an-Donostia, Spain}

 \author{Eduard Carbonel-Sanrom\`{a}}  
   \affiliation{CIC nanoGUNE,  20018 San Sebasti\'an-Donostia, Spain}
 	
 \author{Manuel  Vilas-Varela} 
 	 \affiliation{CIQUS, Universidad de Santiago de Compostela, Spain}

 \author{Aran Garcia-Lekue}
	 \affiliation{Donostia International Physics Center, 20018 Donostia-San Sebastian, Spain}
	% \alsoaffiliation{Ikerbasque, Basque Foundation for Science, Bilbao, Spain}
	 	 \affiliation{Ikerbasque, Basque Foundation for Science, Bilbao, Spain}
 
 \author{Enrique Guiti\'an}   
 	 \affiliation{CIQUS, Universidad de Santiago de Compostela, Spain}

 \author{Diego Pe\~na}
	  \affiliation{CIQUS, Universidad de Santiago de Compostela, Spain}
	  \email{diego.pena@usc.es}
  
 \author{Jose Ignacio Pascual} 
 	  \affiliation{CIC nanoGUNE,  20018 San Sebasti\'an-Donostia, Spain}
 	  	 \affiliation{Ikerbasque, Basque Foundation for Science, Bilbao, Spain}
		 %\alsoaffiliation{Ikerbasque, Basque Foundation for Science, Bilbao, Spain}
	\email{ji.pascual@nanogune.eu}

\begin{abstract}
    %\setstretch{1.5} 
	\textbf{Large aromatic carbon nanostructures are cornerstone materials due to their increasingly active role in functional devices, but their synthesis in solution encounters size and shape limitations. New on-surface strategies facilitate the synthesis of large and insoluble planar systems with atomic-scale precision. While dehydrogenation is usually the chemical zipping reaction building up large aromatic carbon structures, mostly benzenoid  structures are being produced.  Here, we report on a new cyclodehydrogenation reaction transforming a sterically stressed precursor with conjoined cove regions   into  planar carbon platform by incorporating azulene moieties in their interior. Submolecular resolution STM is used to characterize this exotic large polycyclic aromatic compound on Au(111) yielding unprecedented insight into a dehydrogenative intramolecular aryl-aryl coupling reaction. The resulting polycyclic aromatic carbon structure shows a   [18]annulene core hosting peculiar pore states confined at the carbon cavity.}
\end{abstract}

\maketitle

\newpage
%   \setstretch{1.5}
%\linenumbers
 
In recent years, solution chemistry has been successfully employed to obtain graphene nano structures  with fascinating structures following a bottom-up approach \cite{JishanWu2007,Pena2010,Chen2012,Segawa2016}. The precise control of their size and shape regarded these carbon nanostructures as technologically relevant materials in electronic and optoelectronic applications, in which large electron mobility and robust structure are stringent requirements \cite{Georgakilas2015,Ferrari2015}.  However,  the large size and the extremely low solubility of these molecules severely complicates their manipulation and characterization by standard methods. Recently, combined sub-molecular atomic force microscopy (AFM) and scanning tunneling microscopy (STM) have  emerged as  methods to study insoluble nanographenes prepared by solution chemistry~\cite{Schuler2014,Rogers2015,Monig2016,Pavlic2017}. 
An important drawback of these methods is that they require  sublimation of the material on an adequate surface, which limits the size of the molecules that can be studied. To overcome this limitation, an alternative approach relies on the preparation by solution chemistry of key building blocks that can be sublimated onto a surface, and act as precursors for the final on-surface preparation of the corresponding nanomaterial \cite{Grill2007,Franc2011,Mendez2011,Lindner2015,Nacci2016}.  This approach is usually based on a sequence of thermally activated Ullmann couplings (UC) to polymerize dibromo-substituted monomers, followed by surface-induced cyclodehydrogenation reactions (CDH). The last CDH step is assisted by the  metal surface to planarize the reaction products, turning molecular sites with high steric hindrance into benzenoid rings \cite{Treier2011}. This protocol have been extensively used for the on-surface preparation of graphene nanoribbons (GNRs)~\cite{Cai2010,Narita2015}.

 \begin{figure*}  [ht]
 	\centering
 	\includegraphics[width=0.75\textwidth]{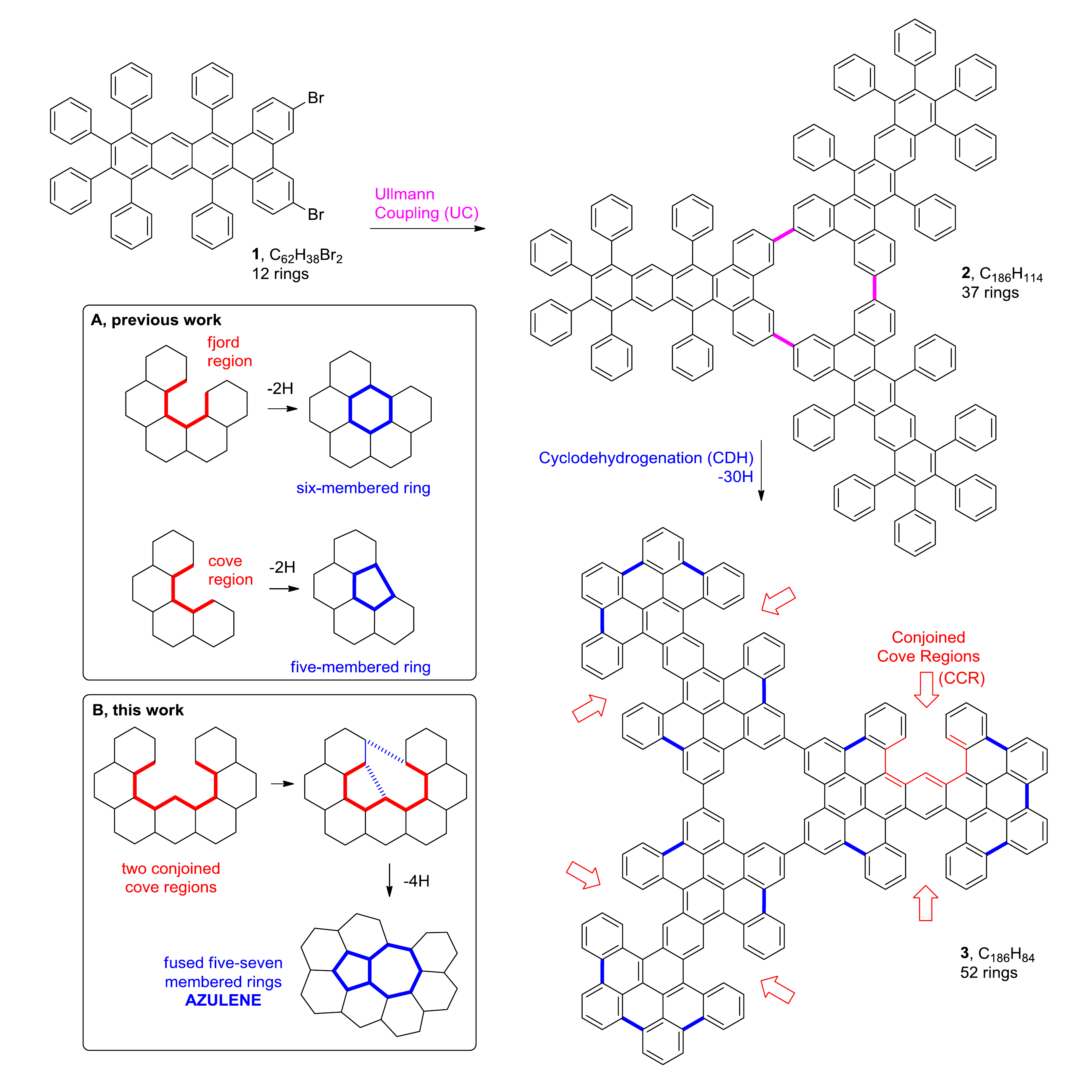}    	
 	\caption{ \textbf{Bottom-up strategy to fabricate large carbon nanostructures:}  Chemical structure of the molecular precursor \textbf{1} synthesized by aryne cycloadditions (see Supporting Note 1). Structure \textbf{2} is the expected nanographene after Ullmann coupling three precursors \textbf{1}, to form an [18]annulene core, while structure \textbf{3} is the expected product after cyclodehydrogenation of the peripheral phenyls. Red arrows indicate the conjoined cove regions with high steric repulsion, described later in the manuscript. The new bonds after each step are highlighted in pink (UC) and in blue (CDH). Square A shows previous C-C zipping CDH reactions reported on a metal surface, creating six-membered rings \cite{Treier2011} and five-membered rings \cite{Otero2008} from  fjord and cove regions, respectively. Square B shows the reported CDH reaction starting from a conjoined cove region, resulting in the formation of azulene moieties (dashed blue lines show new bonds).  \label{Chemstructure}}
 \end{figure*}

Based on this strategy, we aimed at combining UC reactions with a CDH cascade to produce large nanographene structures on a surface.  In particular, we strived for covalently coupling three graphene blades into a trimer-like shape with a central [18]annulene pore (structure \textbf{3} in Figure 1)   using  the 12-ring dibromo polycyclic aromatic compound \textbf{1} as precursor. 
The [18]annulene pore is the smallest possible that can be created by means of C-C coupling strategies~\cite{Fan2013,Chen2016}, and the prediction is that it can host peculiar electronic features confined at the pore~\cite{Zhang2016}. We found that the resulting structure \textbf{3} suffers strong intramolecular torsion at the carbon blades due to the presence of six conjoined cove regions with high steric hindrance. The structure spontaneously transforms into a new type of planar nanographene incorporating six azulene moieties (i.e. a seven-membered ring fused to a five-membered ring). 

The structure of the produced nanographenes implies the discovery of  a novel CDH reaction: while it is well documented  the CDH of \textit{fjord} cavities to form six-membered rings \cite{Treier2011}, as well as the CDH of \textit{cove} regions to afford pentagons \cite{Otero2008} (Fig. 1a), here we describe the CDH of \textit{conjoined cove}  regions  (CCR) to obtain azulene fragments (Fig. 1a). Notably, since the azulene motif is the basic distortion of a graphenoid structure, named as Stone-Wales (SW) defect~\cite{Terrones2000,Ma2009}, this novel CDH of conjoined cove regions provides a synthetic tool to introduce SW defects in nanographene structures.

%\section{Results \& discussion} 

\begin{figure*} [t]
	\centering
	\includegraphics[width=0.75\textwidth]{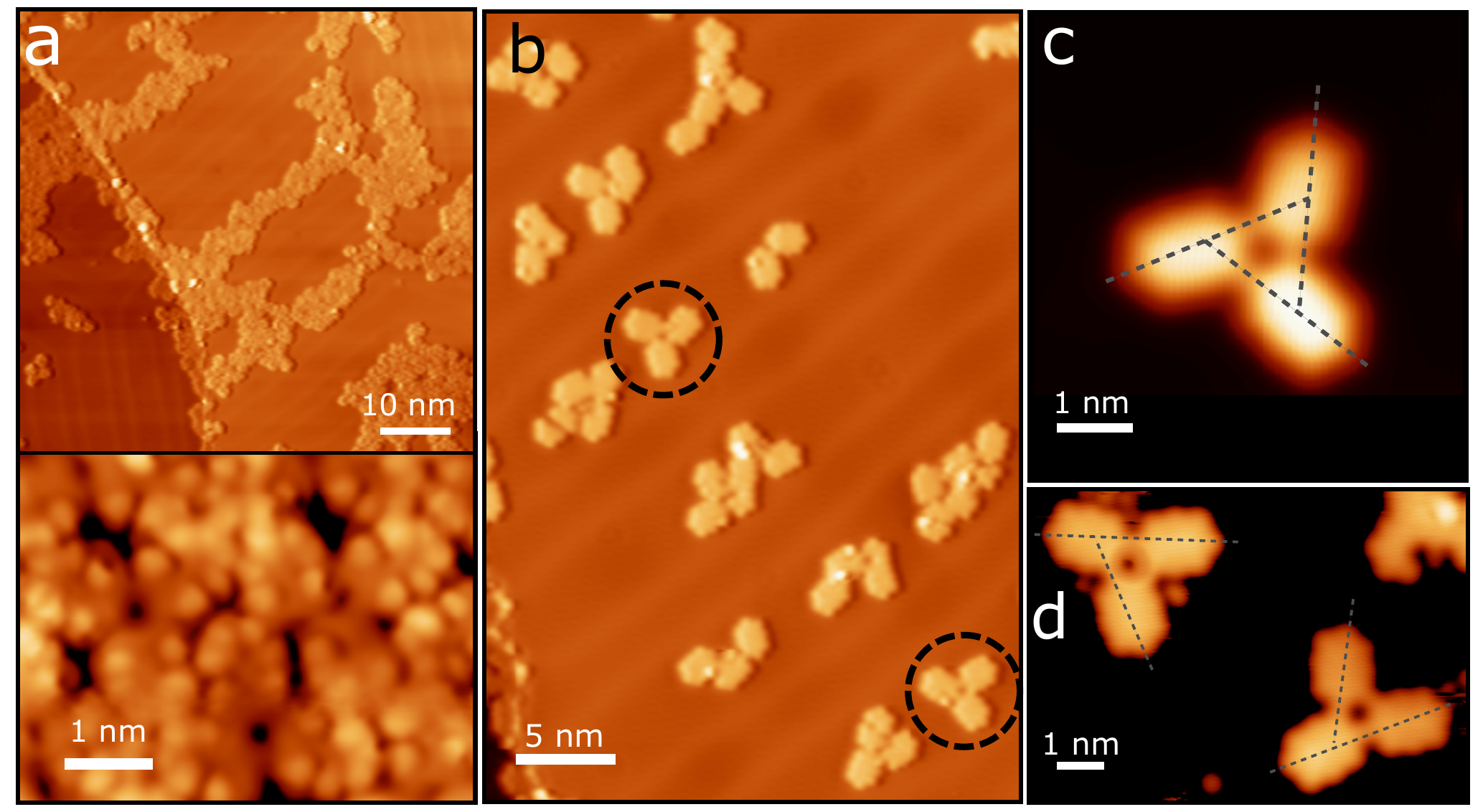} 
	\caption{\textbf{On surface synthesis of large carbon nanostructures:} Large scale STM topography image of the reacted monomers \textbf{1}, arranged in heterogeneous $\pi$-stacked islands. Inset shows a small scale STM topography. Individual monomer  \textbf{1} appear interconnected, showing protrusions coinciding with the positions of peripherial phenyl substituents in \textbf{2}, suggesting  that the molecules did not undergo a CDH at this temperature  (\vs=0.68 V, \It=0.12 nA). 	
	(b) Large scale STM topography image of the cyclodehydrogenated phase, characterized by the small disperse arrangements of molecules. Dashed circles highlight trimer structures (\vs=1.0 V, \It=0.3 nA).  (c) Small scale constant height image of a propeller shape C3-symmetric trimer, the three blades share the same relative directionality (gray dashed lines) (\vs=1.2  V, \It=0.5 nA).  (d) Small scale topography image of two asymmetric trimers. Dashed lines indicate the axis of each blade (\vs=1.0 V, \It=0.5 nA).   \label{STMTopo}}
\end{figure*}

The on-surface process followed two steps: first, a sub-monolayer amount of precursor \textbf{1} was sublimated under ultra-high vacuum conditions  onto a clean  Au(111) substrate (see Methods). During deposition, the metal sample was held at (200 $^{\circ}$C)   to  favour the  Ullmann-coupling polymerization of \textbf{1} upon  arrival to the surface,  rather than their packing in more complex $\pi$-stacked ensembles. Still, STM inspection of the sample  after this intermediate step  finds that the reacted monomers formed closed-packed islands of different sizes and shapes (Figure~\ref{STMTopo}a) but did not undergo CDH yet~\cite{Treier2011}.   Second,  the precovered sample (as in  Fig.~\ref{STMTopo}a) was annealed to temperatures around 375 $^{\circ}$C. This converted the packed molecular islands into a disperse set of clusters of interconnected structures (Fig.~\ref{STMTopo}b). The flat appearance of the structures in the STM images suggests that most of the  polymers underwent a substrate-induced cyclodehydrogenation, which fused all their phenyl moieties into a graphenoid carbon structure  (a very small amount of unreacted structures can still be distinguished). %The planar shape of such new structures favours the interaction with the metal surface instead of intermolecular $\pi$-stacking and, consequently, the reacted species appear arranged in small disperse structures rather than in packed molecular islands. 

%as we show later, structure \textbf{3} is not planar, experiencing a strong twist caused by the steric  repulsion of five hydrogen atoms in the  \mbox{CC} regions, Fig.~\ref{Chemstructure}. 

A fraction  of the reacted polymers corresponds to structures with a trimer-like shape, composed of three flat blades around a central pore, which we attribute to the [18]annulene core (Figure \ref{STMTopo}c).  These shapes resemble the expected product  of the two step-reaction, sequentially producing 3 C-C bonds by Ullmann coupling and 15 C-C bonds by intramolecular cyclodehydrogenation. However,  a closer inspection reveals that although these trimers are flat they deviate from  structure \textbf{3}: they show a faint helical shape endowed by the small bend of all  three blades  with  the same  orientation with respect to the pore. Additionally, we observe other trimers even deviating from the C3-symmetric forms, in which  one of the blades bends  opposite  to the others   (Figure \ref{STMTopo}d). We did not find in our experiments any trimer with shape  resembling \textbf{3}, but only with (C3-symmetric)  propeller shape  and asymmetric trimers, as in Figs. \ref{STMTopo}c and \ref{STMTopo}d. The observation of these unexpected shapes suggested that an alternative cyclization process occurred during  the reaction. 

%\subsection{High-resolution STM imaging:} 

\begin{figure*}[th]
 	\centering
 	\includegraphics[width=0.7\textwidth]{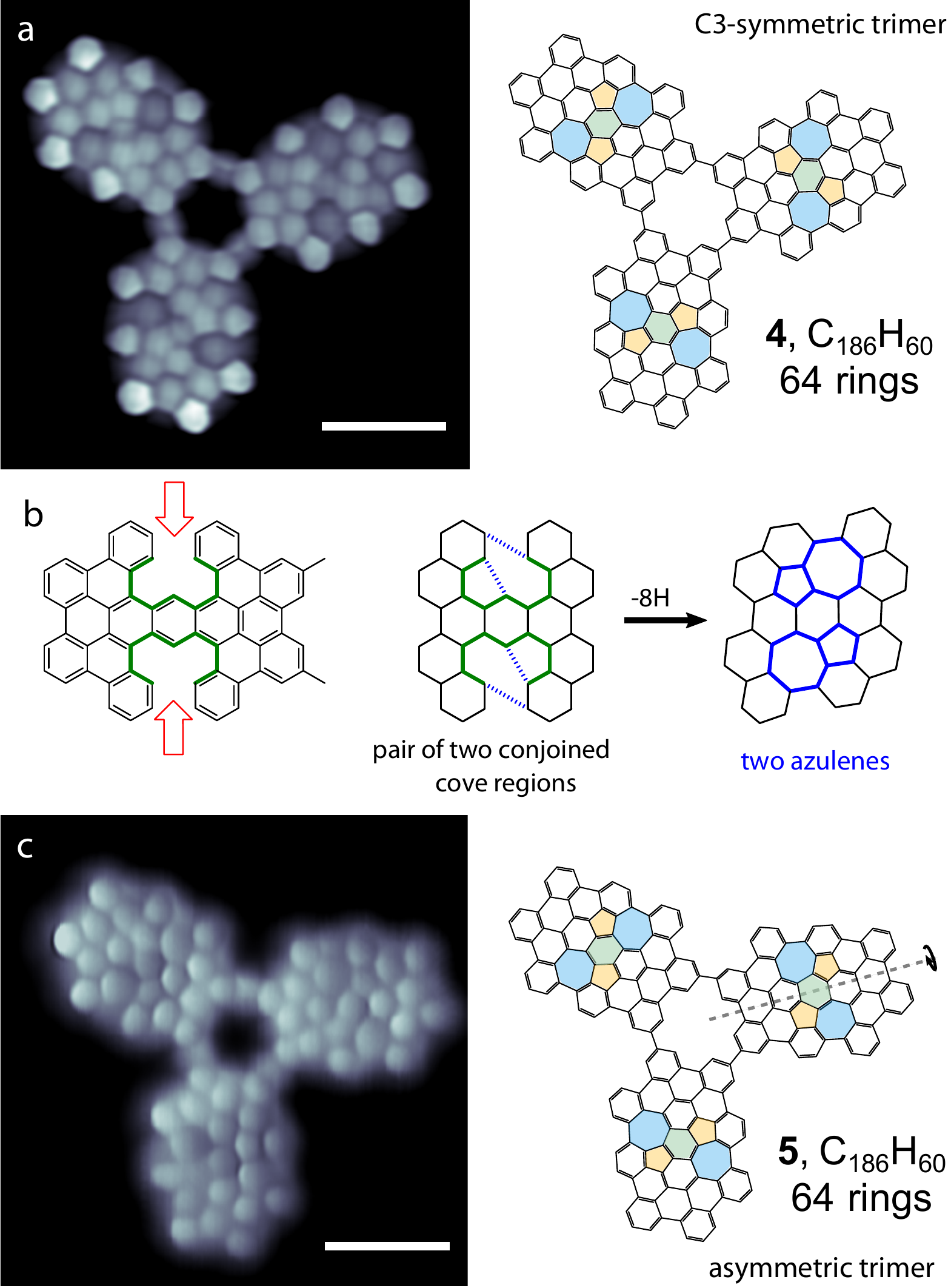} 
 	\caption{\textbf{High resolution STM images revealing a new flattening reaction:} (a) Constant height \didv\, map with a CO functionalized tip of a propeller shape C3-symmetric trimer (structure \textbf{4})  showing its  ring structure. Each of the three blades possesses   two azulenes fused to a benzene ring (highlighted in colour in \textbf{4}), which share the same orientation respect to the centre of the trimer. (b) Scheme showing the formation of two azulene moieties from  two conjoined cove regions in each of the blades of \textbf{3}, with new bonds indicated with dashed lines.  (c) Same as in panel \textit{a}, but for an asymmetric trimer \textbf{5}. The high resolution image evidences that one of the blades  is reversed with respect to \textbf{4},  which is the cause of the asymmetry on the trimer. Scale bars are 1 nm (\vs=5~mV, \It=1~nA, scale bar: 1.2~nm).  \label{COSTM}}
 \end{figure*}

To find out the origin of the asymmetric shapes, we resolved the intramolecular structure of the trimers with high-resolution STM images using a CO functionalized tip (see Methods). The presence of a CO molecule at the tip apex allows approaching the tip to the onset of repulsive Pauli forces  and transduce  variations of atomic-scale forces into changes of tunneling current and differential conductance \didv~\cite{Temirov2008,Weiss2009,Gross2010,Hapala2014,Jarvis2015} with high resolution. In figure~\ref{COSTM}a, we show a constant height \didv\, map of a C3-symmetric trimer, resolving with high detail its intramolecular ring structure. Each of the three planar blades is composed of 21 rings instead of the expected 17 rings in \textbf{3}. Hence,  the resulting trimers \textbf{4}  possess in total 64 carbon rings, 12 more than compound \textbf{3}. 
Furthermore, the  conjoined cove regions of structure  \textbf{3} are missing and, instead, two azulene moieties appear in each of the three blades.

Five- and seven-membered  rings are typically observed as defects in the graphene structure~\cite{Cockayne2011,Rasool2011,Banhart2011,Biro2013,Yazyev2014,Liu2016}, and appear commonly at the connecting boundaries between two graphene sheets with different alignment~\cite{Dienel2015}. They are also allotropic forms of graphene, which maintain the $sp^2$ character of the carbon sheet. In the current case, the two azulene moieties appearing  inside each of the three planar blades show a well-defined and repetitive configuration, which refrain us from calling them defects. They always appear forming   a perylene isomeric motif, where two azulenes are fused to the same benzene ring as  depicted in Fig. 3b. We are not aware of previous synthesis strategies to produce this moiety, which turns out to endow flexibility to flat graphenoid sheets~\cite{Terrones2000,Ma2009}. 

The resulting azulene moieties are the origin of the propeller shape and asymmetric  structure of the trimers. They show two possible (mirror-symmetric) orientations with respect  to the axis of the carbon blade (the radius of the trimer), imposing the bending distortion apparent already in Figs. \ref{STMTopo}c and \ref{STMTopo}d. 
Three equally oriented azulene pairs in each blade lead to propeller-shaped C3-symmetric trimers \textbf{4} (Figure \ref{COSTM}a), while a
reversed orientation in one of the trimer blades affords  the asymmetric structure \textbf{5} shown in Figure \ref{COSTM}c. 
After STM inspection of the structure of 127 trimers, we found  27  species of the C3-symmetric form \textbf{4}, while the rest are  asymmetric trimers \textbf{5}. The ratio of \textbf{4}  vs. \textbf{5} species  is close to the 1 to 3  proportion of possible configurations of each type. Therefore, the formation of azulene moieties in the core of the blades is  not affected by the orientation of the neighbour blades in the trimer, indicating a non cooperative reaction.   Furthermore, the CDH reaction does not require the presence of contacting blades to succeed, since similar azulene moieties were found   in  monomers \textbf{1} as well as in dimers (see Supporting Note 3). 

It is interesting to note that the 7-membered rings are imaged darker than the others, whereas the peripheral benzene rings appear brighter. Ring-specific current contrast appears to be linked to the Clar  sextet structure of graphenoid systems \cite{Dienel2015}.  The structure \textbf{4} accommodates  nine Clar  sextets in each blade, but  only the six outer rings appear brighter in the images, which are the ones  with  probably less distortion.  As we show in the Supporting  Note 4, the different contrast can be well compared with the  electrostatic potential landscape of the nanostructure \cite{Vijayalakshmi2010,Hapala2016}, appearing brighter in the images those rings with larger electron accumulation.  This supports the association of ring contrast in constant-height current images at V$\sim$E$_F/e$ to variations of electron density due to conjugation (as e.g. in Clar  sextets) although other factors may also  affect the tunneling transmission, such as vertical displacements and local density of states around the Fermi level. 
 
%\subsection{On-surface synthesis:} 

 \begin{figure*} [t]
	\centering
	\includegraphics[width=0.75\textwidth]{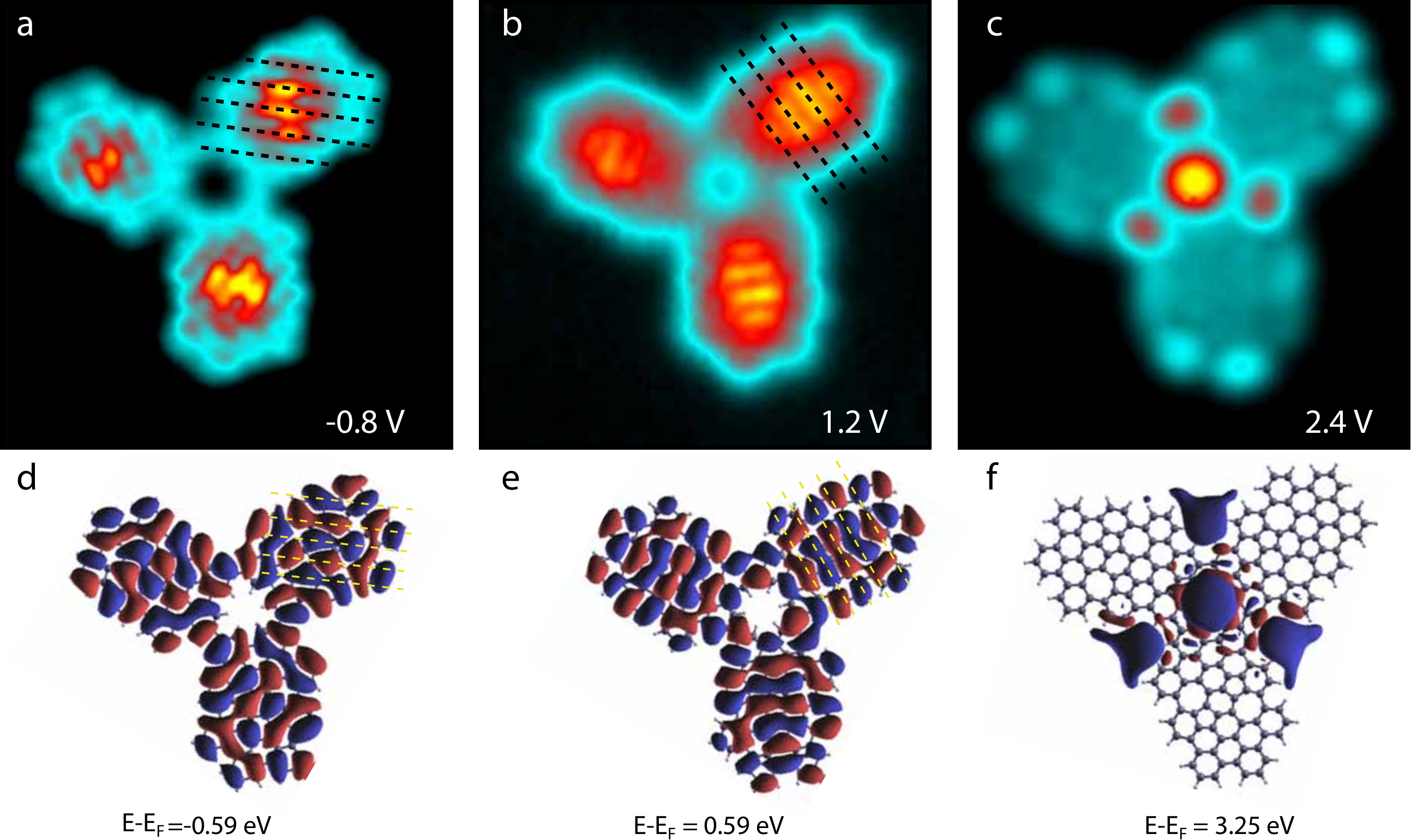}%\\
	\caption{\textbf{STS results and pore state of trimer \textbf{4}:} Constant height \didv\, maps picturing the shape of (a) the HOMO  and (b) LUMO levels  of an propeller shape C3-symmetric trimer \textbf{4}.  (c) Constant height \didv\, maps at higher energy, picturing a molecular state with large weight over the [18]annulene core (see also Supporting Note 4).   		
		(d,e,f) Isosurfaces of constant wave-function amplitude and phase (colour) obtained from DFT simulations of the free  trimer \textbf{4}.  The orientation and number of modes of HOMO and LUMO orbitals (dashed black lines), as well as the shape of the pore state, shows remarkable agreement with the experimental maps (a:\It=0.1~nA; b:\It=0.07~nA; c:\It=0.07~nA).  \label{STSDFT}} 
\end{figure*}

%Since the formation of azulene rings apparently involves  heavy distortions of the precursor  molecule, we analyze the formation pathways using Force-field simulations (see Methods).  
The appearance of azulene moieties plays a key role in the planarization of the molecule. Compound \textbf{3} is heavily distorted in gas phase due to the overlap of five hydrogen atoms at each  conjoined cove region, three of them lying directly over the same site. The steric repulsion induces an out-of-plane twist in the free  compound \textbf{3} ($\sim$ 25 degrees, see results of Force-Field simulations in Supporting Note 2), which hinders the planar adsorption on the surface.  Compound \textbf{4} is on the contrary rather planar, and the azulene fragments simply endow the  three blades  some more flexibility than pure graphenoid structures~\cite{Terrones2000,Ma2009}.  
On a metal surface,  the twisted structures \textbf{3} are forced to planarize by the combined effect of van der Waals forces  and intramolecular restoring forces, both acting against  steric hindrance between H atoms  and weakening the C-H bonds. The catalytical role of the metal substrate at high temperatures results in the hydrogen removal~\cite{Treier2011}.

Contrary to dehydrogenation reactions of  \textit{fjord}~\cite{Treier2011} and \textit{cove}~\cite{Otero2008} cavities,  the CDH of conjoined cove regions    does not directly lead  to cyclization, but additional molecular twists are required to create new C-C bonds. The deprotonated  species is flat, flexible,   thermally excited,  and active to capture gold adatoms~\cite{Yang2014},  which could facilitate the additional C-C coupling for azulene formation~\cite{Kocic2016}. 
The energy gain of the reaction is clear: the incorporation of azulene moieties in the free molecule reduces the steric energy mostly due to the removal of the strong twist of compound \textbf{3} (Supporting Note 2). On the surface, the energy gain is considerably larger as the more planar structure \textbf{4} maximizes the contact with the gold surface.

%\subsection{Electronic structure:} The azulene moieties in the blades also contribute to the stabilization of the extended aromatic system. 
We investigated the electronic structure of the produced trimers by means of differential conductance (\didv) imaging, exploring the extension of  frontier orbitals. Point spectra acquired over the molecules (shown in the Supporting Note 3) found \didv spectral features at  -0.8 V and 1.2 V associated with HOMO and LUMO derived molecular states.   
Constant height  maps of \didv\  measured at these bias voltage values are shown in Figure \ref{STSDFT}a and  \ref{STSDFT}b, respectively. The \didv\  images show a characteristic intramolecular pattern extending all around the carbon platform rather than localized in some parts or edges, supporting the conjugated character of these states. To interpret these maps, we compared them with   DFT simulations of the density of states (DOS) of the HOMO (Figure \ref{STSDFT}d) and LUMO (Figure \ref{STSDFT}e) states for a fully relaxed,  C3-symmetric trimer \textbf{4} in the gas phase (see Supporting Note  4).  The simulations confirm that the dI/dV modulations in both cases are correlated with the lobes of the orbital wavefunction, which run as axial-wavefronts for the LUMO and tilted-waves for the HOMO.  %The asymmetric trimers show a very similar \didv\ maps for the LUMO, while the HOMO shows a mirror symmetric wave-pattern in the miss-oriented lobe. 

Remarkably, a highly localized conductance feature emerges at the center of the [18]annulene core when the sample bias is tuned to e.g.  2.4 V (Figure \ref{STSDFT}c). The state appears with a peculiar three fold shape and with very little DOS weight in the rest of the carbon platform. Pore-localized states in non-covalent molecular architectures on surfaces have been detected due to the confinement of surface electrons~\cite{Pennec2007,Lobo-Checa2009,Klappenberger2009,Krenner2013}.  The  [18]annulene core is, on the contrary, a rigid structure arising from  covalently bonded precursors.  Our DFT calculations on the free species nicely reproduce the shape of this state (Figure \ref{STSDFT}f), including three smaller maxima at each of the three cavities at the joints between the blades. The pore state is thus an orbital of the trimer structure and can probably be depicted as a symmetry-adapted overlap of three specific blade orbitals \cite{Umbach2013,Zhang2016}. In fact, a precursor state  can be already  imaged for dimer structures appearing occasionally over the sample (see Supporting Note 3).

Interestingly, we find that the pore state belongs to the family of the so-called Super-Atom  Molecular Orbitals (SAMOs \cite{Feng2008}), representing the set of high-lying  unoccupied orbitals with principal quantum number n$\geq$3  \cite{Zhao2009}.  SAMOs have very little weight around the carbon cores, and extend well outside   the molecular backbone, occupying the empty space in cavities and pores~\cite{Zhao2009,Hu2010}.  Due to their large extension, SAMO-states are able to overlap with neighbours to form electronic bands in   molecular crystals.  In fact, the pore state shown in Fig. \ref{STSDFT}c is the result of the in-phase combination of  three n=3  orbitals of the propeller blade,   resulting in a tubular state crossing through the [18]annulene core  (see Supporting Note 4).  

% \section*{Conclusions}
In conclusion, we have shown the on-surface synthesis of a unique propeller-shape nanographene molecule with a [18]annulene pore and the remarkable formation of six azulene moieties at its three blades. The creation of azulene moieties follows a novel cyclodehydrogenation pattern in conjoined cove regions, which leads to two new C-C bonds and relaxation of the twisted regions into a flat-lying molecule on the surface. Our work thus highlights an additional reaction pathways allowed by the on-surface synthesis that can be exploited by clever precursor design, opening new possibilities in the growth of graphene-like structures. For example, the discovered reaction pattern suggests a control strategy to introduce Stone-Wales defects in graphene flakes or ribbons, as well as the synthesis of Haeckelites, which are proposed to endow metal character to flat $sp^2$ carbon flakes \cite{Terrones2000}.  Furthermore, we found  peculiar electronic states residing at the central pore of the trimers at high energies, which could be identified as  SAMO-like resonances of the nanographene with the support of DFT simulations. These image-like states are expected to  endow  porous graphene with novel properties such as new electronic bands or enhance the photochemical reactivity of captured molecular guests.  
 
 \textbf{Organic synthesis:} \textbf{} of the organic synthesis of  dibromo polycyclic aromatic compound \textbf{1}  are given in the Supporting Note1. 
 
\textbf{STM measurements:}
 The experiments were conducted in custom-made Low-Temperature (4.8~K) STM, under Ultra-High Vaccum conditions. We used a  Au(111) single crystal as substrate, prepared by repetitive cycles of Ar$^+$ sputtering and annealing to 500$^\circ$C, until the surface is atomically clean and flat. High resolution STM imaging was done employed CO-functionalized tips.  For this purpose, small amounts of CO were dosed  after the last reaction step on a sample.  Analysis of STM and STS data was performed with the WsXm~\cite{wsxm} and SpectraFox~\cite{Spectrafox} (http://www.spectrafox.com) software packages. 
 
\textbf{Computational details:}
The electronic structure and geometry of compound \textbf{4}, in its free standing form, were calculated
using density functional theory, as implemented in the SIESTA code \cite{Siesta}. All the atoms of the
molecule were fully relaxed until forces were $<$~0.01eV/\AA, and the dispersion interactions were
taken into account by the non-local optB88-vdW functional \cite{vdWDFT}.The basis set consisted of doublezeta
plus polarization (DZP) orbitals and diffuse 3s and 3p orbitals for C atoms, and DZP orbitals
for H atoms. A cutoff of 300 Ry was used for the real-space grid integrations, and the Γ-point
approximation for sampling the three-dimensional Brillouin zone.

\textbf{Acknowledgements:}
 We are indebted to Martina Corso, Dolores P\'erez, and Daniel S\'anchez Portal for fruitful discussions. We acknowledge financial support from Spanish AEI (Collaborative Project MAT2016-78293),  the Basque Government (Dep. Industry, Grant PI-2015-1-42, Dep. Education, Grant PI-2016-1-27), and the Diputacion Foral de Gipuzkoa (Grant No 64/15),the EU project PAMS (610446), the Xunta de Galicia (Centro singular de investigación de Galicia accreditation 2016-2019, ED431G/09), and the European Regional Development Fund (ERDF).

%  \setstretch{1} 
 
\setstretch{1.0} 
 \bibliography{chem-bib}
 \newpage
  
 %  \setstretch{1} 
  %\singlespacing

\end{document}

\section{Supporting Notes for the manuscript: \\  	On-surface route for producing planar nanographenes with  azulene moieties.}
 
   \vspace{1cm}
\setstretch{1.0} 
%\tableofcontents

    \setstretch{1.4}

\addcontentsline{toc}{section}{Note 1: Organic synthesis of the molecular precursor}
\section*{Note 1: Organic synthesis of the molecular precursor} \label{SN-OS}

\subsection{General Methods:}
All reactions were carried out under argon using oven-dried glassware. TLC was performed on Merck silica gel 60 F254; chromatograms were visualized with UV light (254 and 360 nm). Flash column chromatography was performed on Merck silica gel 60 (ASTM 230-400 mesh). $^{1}$H and $^{13}$C NMR spectra were recorded at 300 and 75 MHz (Varian Mercury 300 instrument), respectively. Low-resolution electron impact mass spectra were determined at 70 eV on a HP-5988A instrument. High-resolution mass spectra (HRMS) were obtained on a Micromass Autospec spectrometer.

Molecules \textbf{6}, \textbf{7} and \textbf{8} were prepared following published procedures  \footnote{a) For preparation of molecules \textbf{6} and \textbf{7}, see: H. Wettach \textit{et al.} \textit{J. Mater. Chem.} \textbf{21},  1404-1415 (2011).\newline
b) For synthesis of molecule \textbf{8}, see: H. M. Duong \textit{et al.} \textit{Org. Lett.} \textbf{5},  4433-4436 (2003).
}. Commercial reagents and anhydrous solvents were purchased from ABCR GmbH, Aldrich Chemical Co. or Strem Chemicals Inc., and were used without further purification. \textit{n}-BuLi was used in solution in hexane (2.4 M).

\begin{figure} [!h] 
	\begin{center}
	\includegraphics[width=0.9\textwidth]{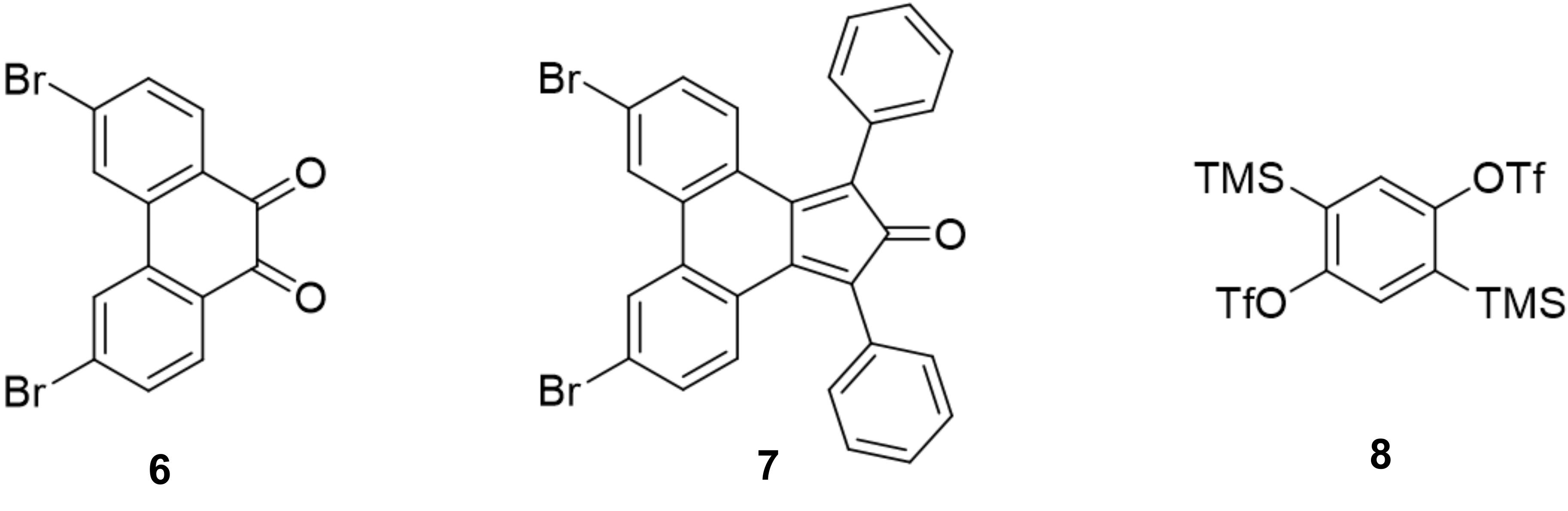}
	\end{center}\label{precursorssup}
		\caption{Chemical structure of intermediate molecules \textbf{6}, \textbf{7} and \textbf{8}. }
	\end{figure}
	
\newpage
	
\subsection{Synthesis of 3,6-dibromophenanthrene-9,10-dione (6):}
To a solution of 9-10-phenantroquinone (\textbf{9}, 10.2 g, 49 mmol) and benzoyl peroxide (0.97 g, 4.00 mmol) in nitrobenzene (75 mL) Br$_{2}$ (5.1 mL, 100 mmol) was added over 30 min. at room temperature. The resulting mixture was heated at 110$^{\circ}$C for 16h. After cooling, the reaction crude was filtered and the residue was washed several times with hexane affording compound \textbf{6} (11.7 g, 66\%) as a yellow solid. $^{1}$H NMR (298 K, 300 MHz, CDCl$_{3}$) $\delta$: 8.10 (d, \textit{J} = 1.6 Hz, 2H), 8.05 (d, \textit{J} = 8.3 Hz, 2H), 7.65 (dd, \textit{J} = 8.3, 1.5 Hz, 2H) ppm.

\begin{figure}  [!h] 
	\begin{center}
    %\vspace{1cm}
	\includegraphics[width=0.6\textwidth]{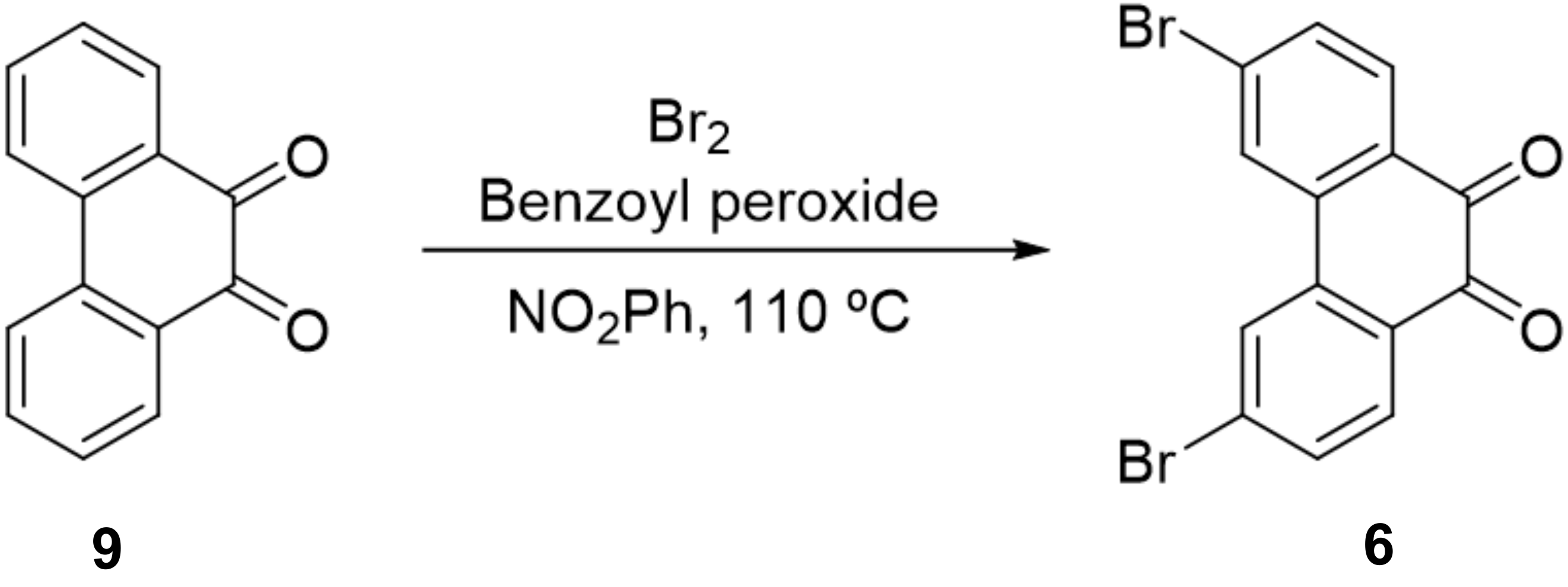}
	\end{center}
			\caption{Synthesis of compound \textbf{6}.}
\end{figure}

\subsection{Synthesis of 6,9-dibromo-1,3-diphenyl-2H-cyclopenta[l]phenanthren-2-one (7):}

A mixture of compound \textbf{6} (1.00 g, 2.75 mmol) and 1,3-diphenylpropan-2-one (625 mg, 2.91 mmol) in MeOH (40 mL) was heated at 80 $^{\circ}$C while a solution of KOH (160 mg, 2.85 mmol) in MeOH (4 mL) was slowly added. After addition, the mixture was refluxed for 30 min. and then cooled in an ice bath. The precipitate formed was filtered and washed with MeOH (4 x 20 mL) to yield compound \textbf{7} (1.13 g, 76\%) as a brownish solid. $^{1}$H NMR (298 K, 300 MHz, CD$_{2}$Cl$_{2}$) $\delta$: 7.93 (d, \textit{J} = 2.0 Hz, 2 H); 7.46-7.32 (m, 12 H); 7.11 (dd, \textit{J} = 1.9 Hz, \textit{J} = 8.6 Hz, 2 H) ppm.

\begin{figure}  [!h] 
	\begin{center}
		\includegraphics[width=0.7\textwidth]{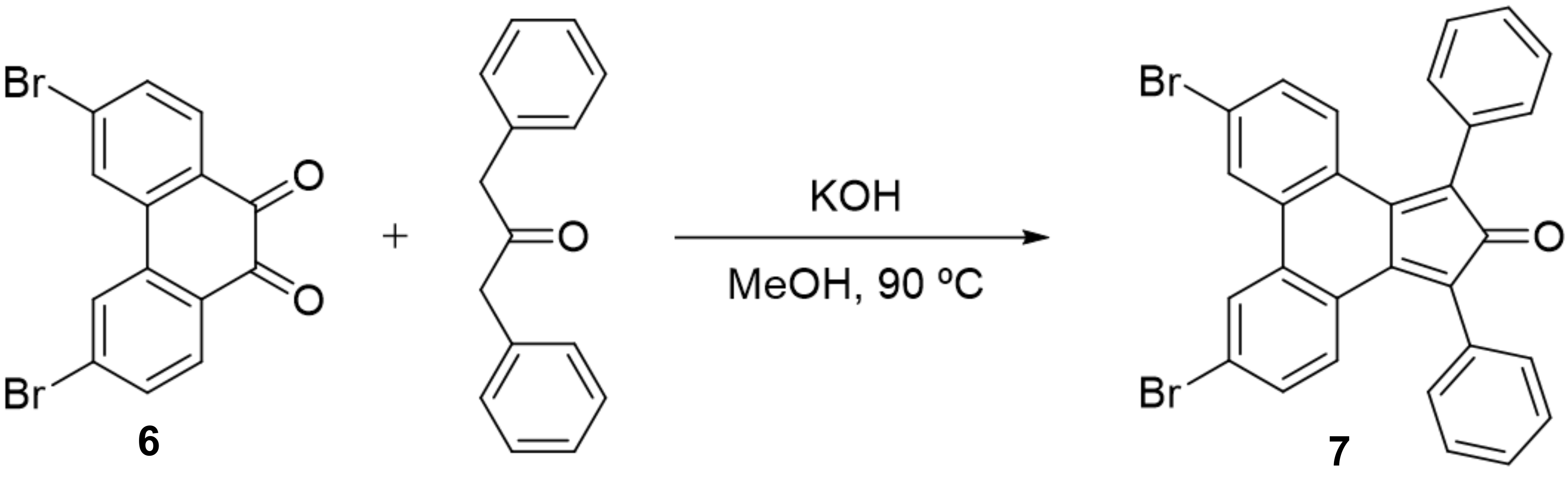}
	\end{center}
		\caption{Synthesis of compound \textbf{7}.}
\end{figure}

\subsection{Synthesis of 3,6-dibromo-9,14-diphenyl-12-(trimethylsilyl)benzo[f]tetraphen-11-yl trifluoro-\\methanesulfonate (10):}

To a mixture of \textbf{7} (200 mg, 0.38 mmol) and \textbf{8} (220 mg, 0.42 mmol) in MeCN/CH$_{2}$Cl$_{2}$ (1:1, 28 mL) anhydrous CsF (84 mg, 0.55 mmol) was added. The resulting mixture was heated at 50$^{\circ}$C for 16 h. Solvents were removed under vacuum and the residue was purified by column chromatography (SiO$_{2}$; hexane/CH$_{2}$Cl$_{2}$ 4:1) to afford \textbf{10} (116 mg, 37\%) as a greenish solid. $^{1}$H-NMR (300 MHz, CDCl$_{3}$) $\delta$: 8.34 (s, 1H), 8.33 (s, 1H), 8.14 (s, 1H), 7.86 (s, 1H), 7.62 – 7.50 (m, 10H), 7.44 (dd, \textit{J} = 9.0, 1.9 Hz, 2H), 7.14 (d, \textit{J} = 8.9 Hz, 2H), 0.35 (s, 9H). ppm. $^{13}$C-NMR (75 MHz, CDCl$_{3}$) $\delta$: 152.9 (C), 140.4 (C), 140.0 (C), 136.9 (CH), 135.9 (C), 135.7 (C), 133.0 (C), 132.7 (C), 132.6 (C), 132.2 (2CH), 132.03 (2CH), 131.97 (CH), 131.4 (C), 130.1 (C), 130.0 (C), 129.9 (C), 129.6 (2CH), 129.5 (2CH), 129.4 (2CH), 129.3 (2CH), 128.7 (C), 128.4 (CH), 128.3 (CH), 126.2 (2CH), 122.0 (C), 121.8 (C), 118.44 (q, \textit{J} = 320 Hz,CF$_{3}$), 115.8 (C), -0.88 (3CH$_{3}$) ppm. MS (EI) \textit{m/z} (\%): 808 (M$^{+}$, 80), 675 (28), 660 (16), 596 (17), 424 (14), 243 (11), 73 (100). HRMS (EI) for C$_{38}$H$_{27}$O$_{3}$F$_{3}$SiSBr$_{2}$, calculated: 805.9769, found: 805.9739.

\begin{figure}[!h] 
	\begin{center}
	\includegraphics[width=0.9\textwidth]{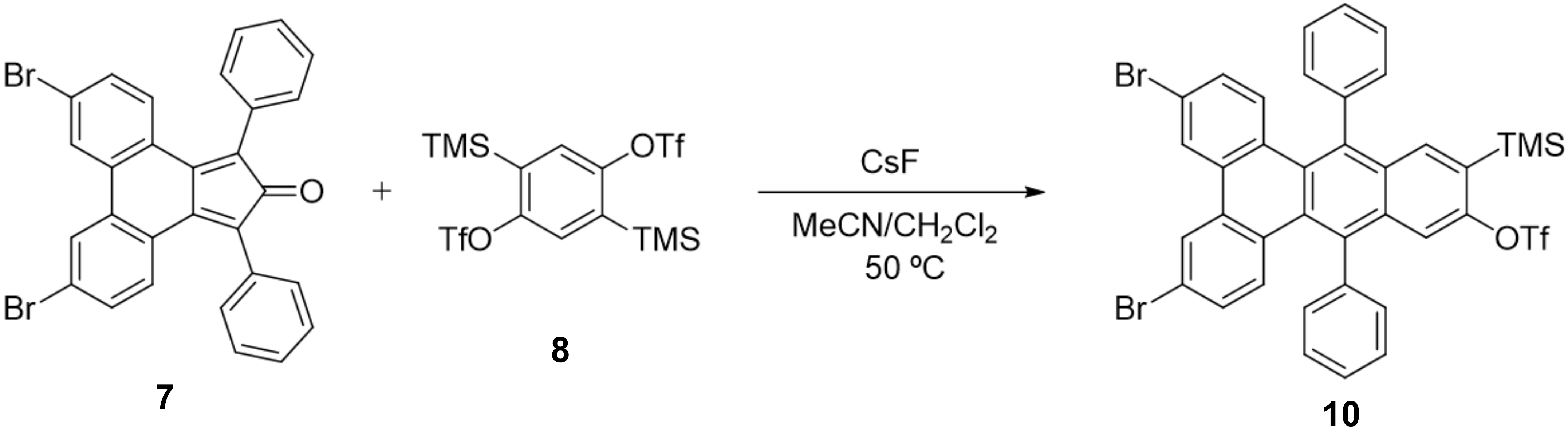}
		\end{center}
	\caption{Synthesis of compound \textbf{10}.}
\end{figure}

\subsection{Synthesis of 3,6-dibromo-9,11,12,13,14,16-hexaphenyldibenzo[a,c]tetracene (1).}
To a mixture of \textbf{10} (75 mg, 0.09 mmol) and tetraphenylcyclopentadienone (30 mg, 0.08 mmol) in MeCN/CH$_{2}$Cl$_{2}$ (1:1, 4 mL), anhydrous CsF (42 mg, 0.28 mmol) was added. The resulting mixture was heated at 50$^{\circ}$C for 16 h. Solvents were removed under vacuum and the residue was purified by column chromatography (SiO$_{2}$; hexane/CH$_{2}$Cl$_{2}$ 3:2) to afford \textbf{1} (45 mg, 61\%) as a yellow solid. $^{1}$H-NMR (300 MHz, CDCl3) $\delta$: 8.21 (s, 2H), 8.20 (s, 2H), 7.37 – 7.29 (m, 12H), 7.17 - 7.10 (m, 10H), 7.04 (dd, \textit{J} = 9.1, 1.5 Hz, 2H), 6.90 - 6.81 (m, 10H) ppm. $^{13}$C-NMR (75 MHz, CDCl$_{3}$) $\delta$: 140.8 (2C), 140.5 (2C), 139.2 (2C), 138.8 (2C), 138.5 (2C), 136.0 (2C), 132.9 (2C), 132.4 (2CH), 132.1 (4CH), 131.3 (4CH), 131.1 (4CH), 131.0 (2C), 130.96 (2C), 130.6 (2C), 129.4 (2CH), 129.0 (4CH),127.5 (2CH), 127.4 (4CH), 127.2 (2C), 126.6 (4CH), 126.2 (2CH), 126.2 (2CH), 125.8 (2CH), 125.4 (2CH), 121.4 (2C) ppm. MS (EI) \textit{m/z} (\%): 941 (M$^{+}$, 100), 863 (20), 569 (8), 414 (40). HRMS (EI) for C$_{62}$H$_{38}$Br$_{2}$, calculated: 940.1340, found: 940.1348.

\begin{figure}  [h] 
	\begin{center}
    	\includegraphics[width=0.8\textwidth]{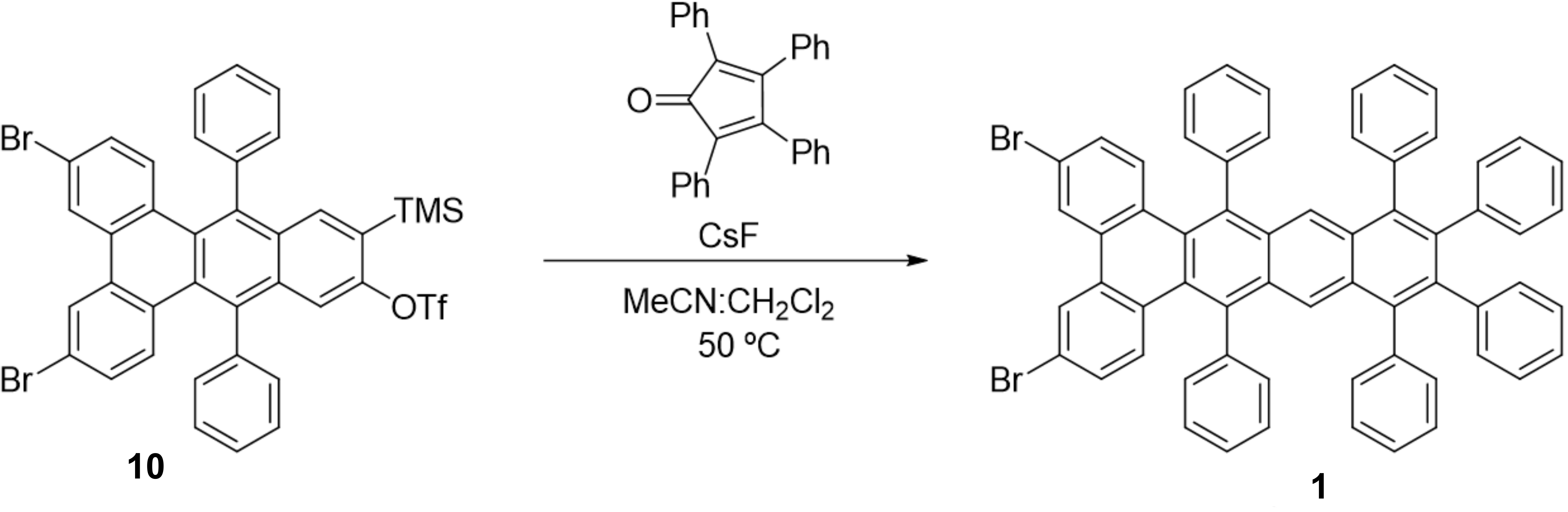}
		\end{center}
			\caption{Synthesis of compound \textbf{1}, the molecular precursor used in this study.}
\end{figure}

\newpage
 
 \subsection{Spectroscopic data}

\begin{figure}[!b]

\begin{tabular}{c}
		
	\includegraphics[width=0.99\textwidth]{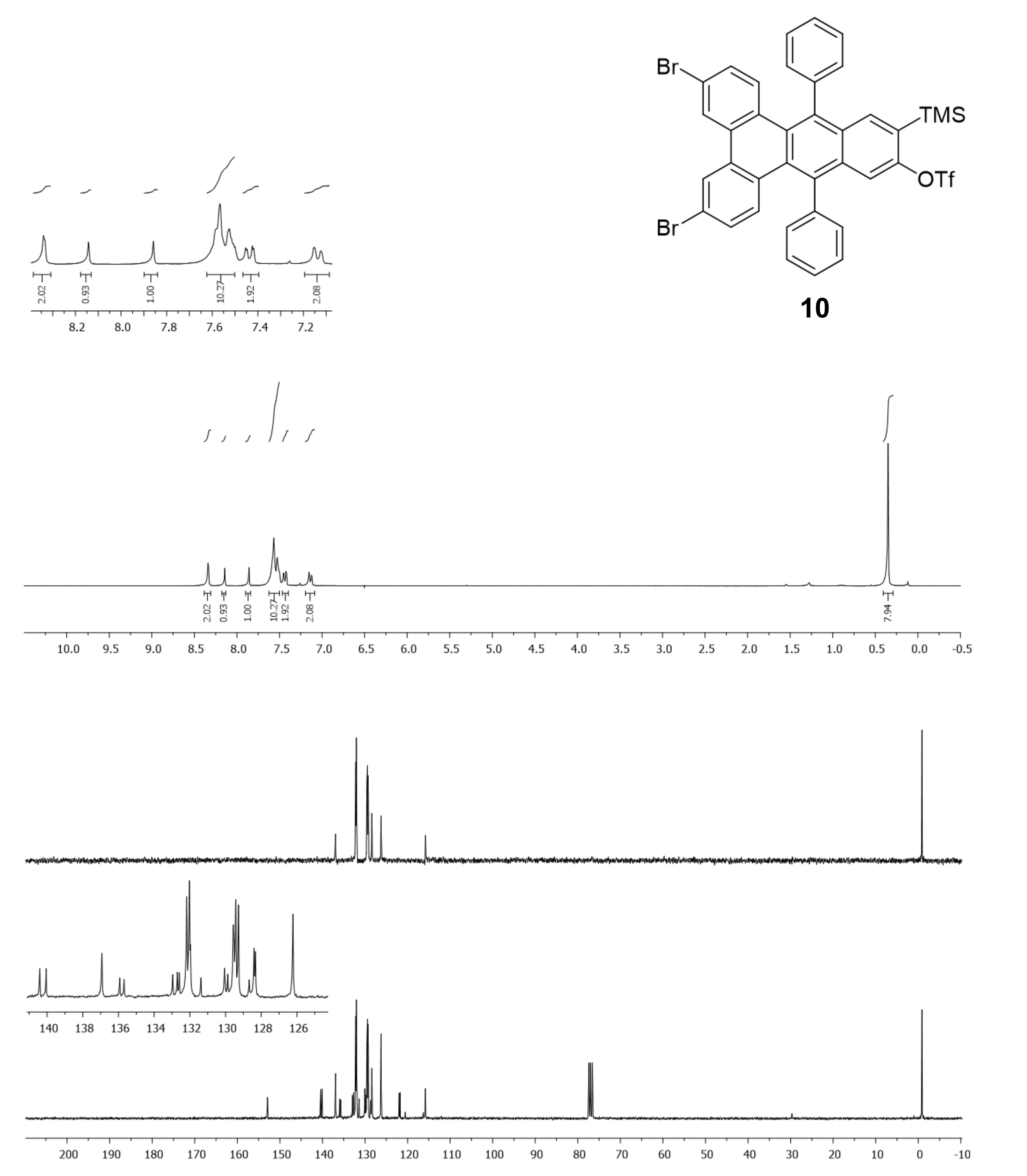} 
		
		\end{tabular}
	\caption{\textsuperscript{1}H and \textsuperscript{13}C NMR spectra of compound   \textbf{10}}  
\end{figure}
 
 \begin{figure}[!t]
  		 	\includegraphics[width=0.99\textwidth]{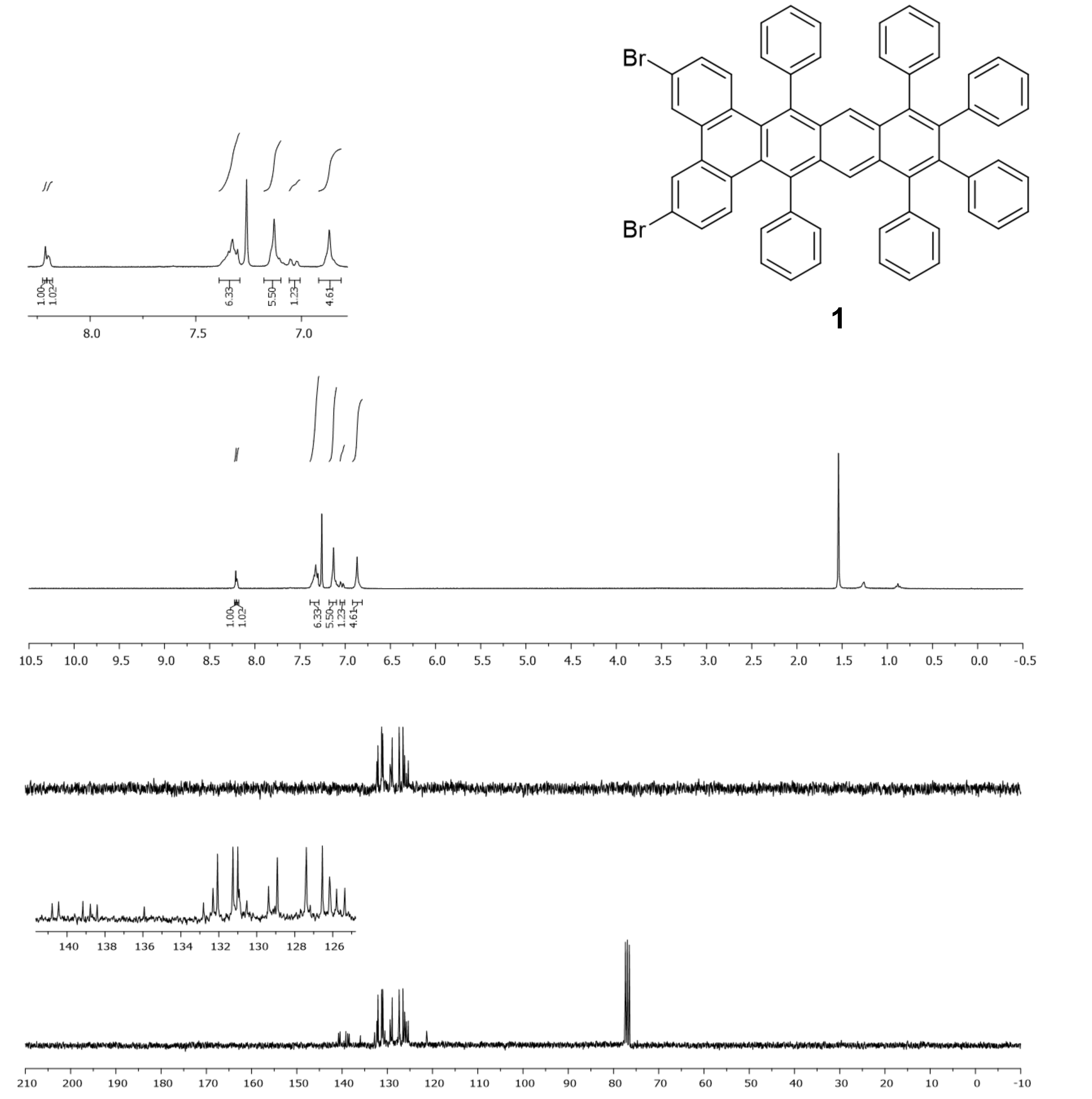}
 	\caption{\textsuperscript{1}H and \textsuperscript{13}C NMR spectra of precursor molecule   \textbf{1}} 
 	\vspace{2cm} 
 \end{figure}

\newpage

\addcontentsline{toc}{section}{Note 2: Molecular structure and steric energy}
\section*{Note 2: Molecular structure and steric energy} \label{SN-react}

\begin{figure} [h]
	\begin{center}
		\includegraphics[width=0.75\textwidth]{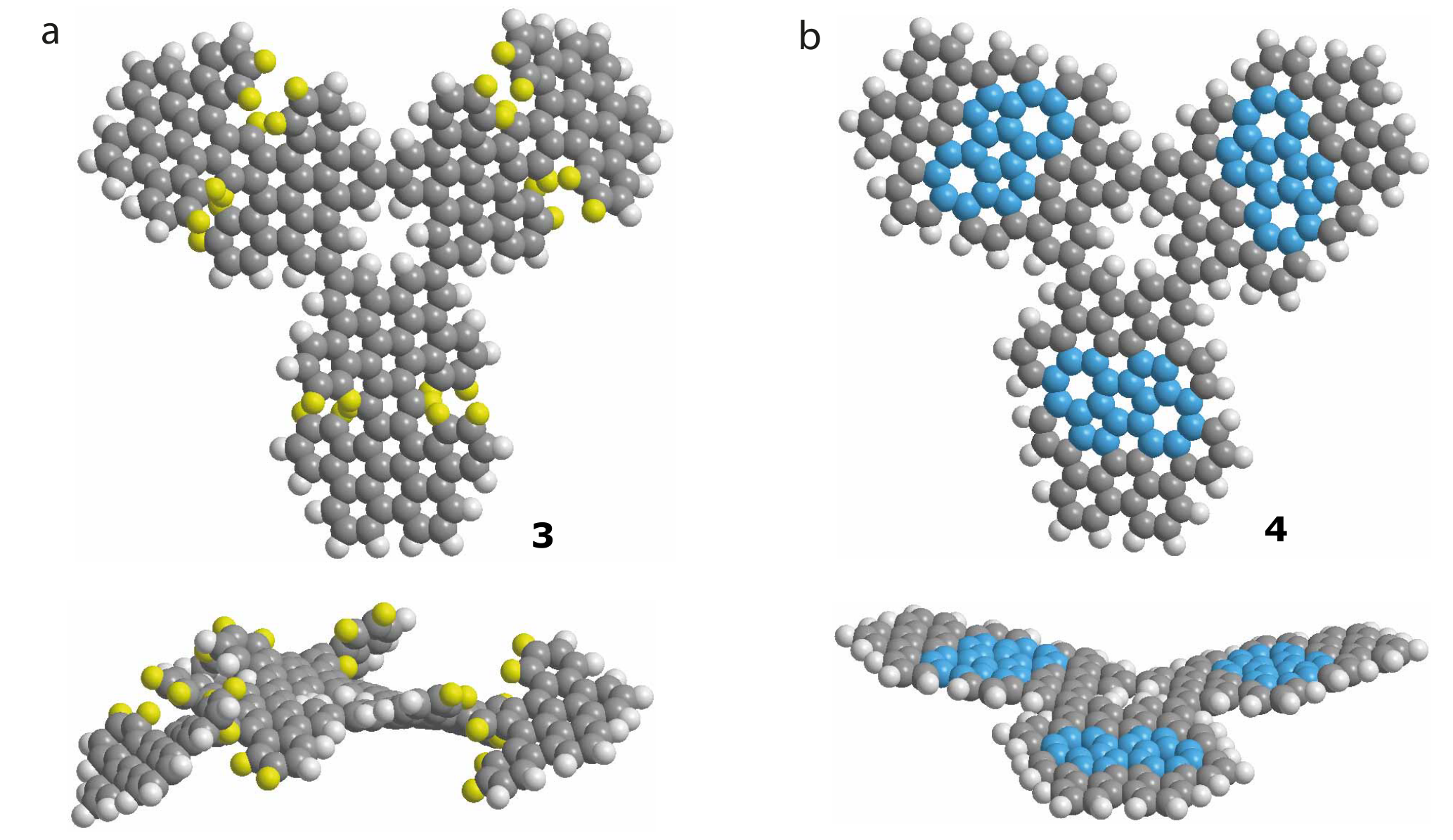}%\\
	\end{center}
	\caption{(a,b) Force Field (MM2) relaxed structures of compounds \textbf{3} and \textbf{4}, respectively.  While \textbf{3} shows a strongly twisted shape, \textbf{4} appears flat after MM2 energy minimization. Yellow atoms in (a) point to the five  H atoms suffering steric repulsion in the CC region. Blue atoms in (b) correspond to the new azulene moiety created after removal of four H atoms from each CC region.  	\label{SN-react1}}
	\end{figure}

Figure  \ref{SN-react1} shows the minimum energy structure of the free compounds  \textbf{3} and \textbf{4} obtained from molecular mechanics simulations using the MM2 Force Field implementation. This method provides the steric energy for each relaxed structure, as a measure of intramolecular distortion from an ideal $sp^2$ force field configuration. We found that both forms have similar steric energy (70 Kcl/mol in 3, 61 Kcal/mol in 4). However, while in \textbf{3} most of the energy comes from dihedral torsions, in \textbf{4} it comes from C-C bond stretch and bends. On a metal surface, planarization forces are expected to reduce the twist angle of  compound \textbf{3}, forcing the hydrogen atoms to approach against their mutual steric hindrance, until dehydrogenation reaction occurs at sufficiently elevated temperatures. The deprotonated radical compound is planar and with negligible steric energy. The formation of compound \textbf{4} from this point requires  in-plane distortions,  probably are mediated by  gold adatoms,  which are prone to bond to active sites in molecular species \cite{Yang2014,Kocic2016}.   

\newpage
\addcontentsline{toc}{section}{Note 3: Scanning Tunneling Spectroscopy results}
\section*{Note 3: Scanning Tunneling Spectroscopy results:}  \label{SN-STS}

\subsection{dI/dV point spectra on trimer (4):}
\begin{figure}[h] 
	\begin{center}
	\includegraphics[width=0.55\textwidth]{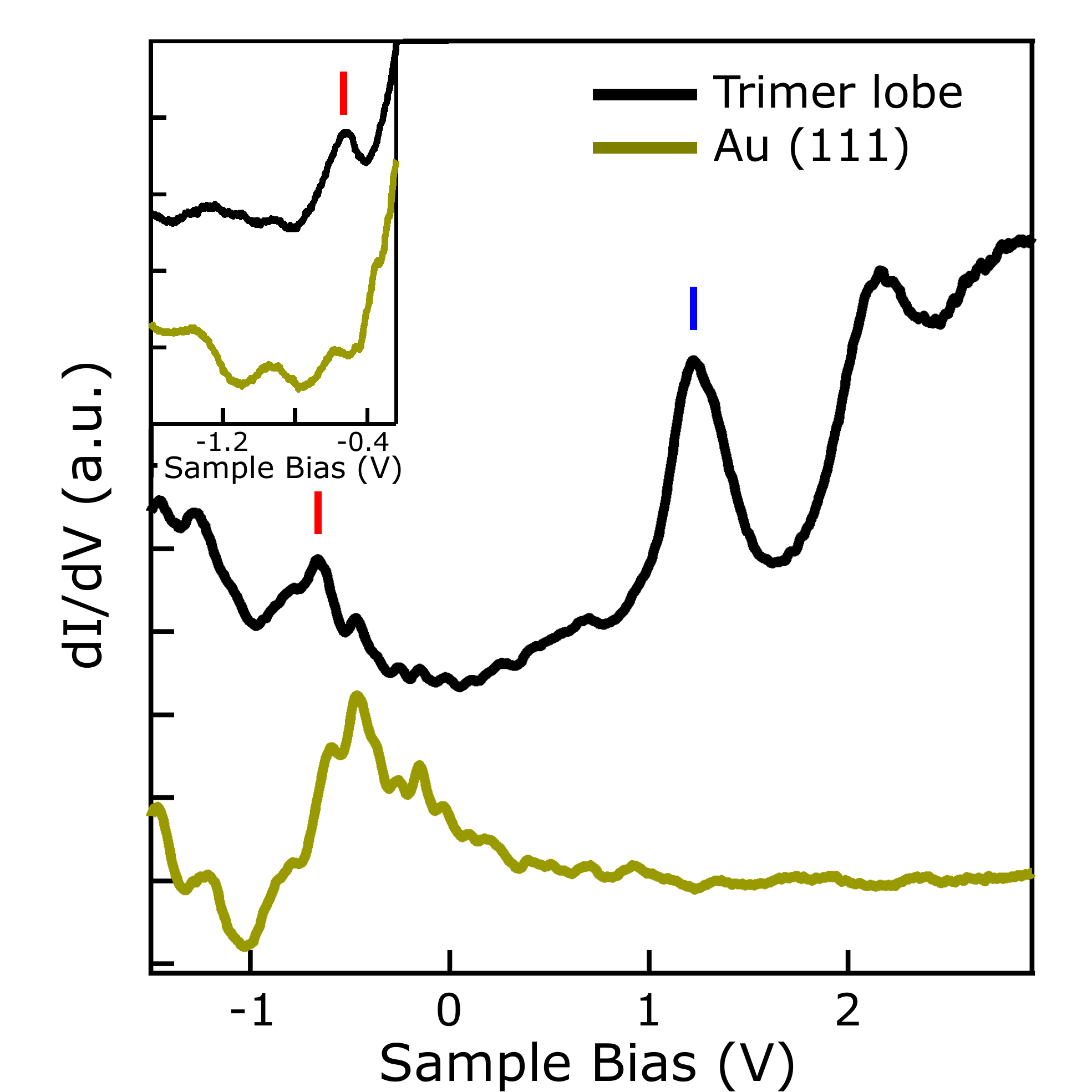}%\\
	\end{center}
	\caption{HOMO and LUMO orbitals of trimer \textbf{4}: (a) dI/dV spectrum taken at the center of a lobe of a trimer (black curve), showing a strong feature at +1.2~V associated to the LUMO orbital of the molecule (\vs=-1.0~V, \It=0.3~nA). The spectrum of the bare Au(111) substrate is shown as a reference (green curve). Inset: constant current spectrum ( \It=0.2~nA) showing a resonance  peak attributed to the HOMO orbitals of the trimer. 
	\label{SN-STS-frontier}}
	\end{figure}

Scanning tunneling spectroscopy was performed over the trimer \textbf{4} to study their electronic configuration. We measured the differential conductance dI/dV  using a lock-in  amplifier. To detect frontier orbitals, we measured constant height spectra on the trimer blades (Fig. \ref{SN-STS-frontier}). The alignment of Highest Occupied and Lowest Unoccupied Molecular Orbitals (HOMO and LUMO) are detected as sudden increase in dI/dV signal at the corresponding bias alignment: -0.7 V and +1.2 V. The constant height dI/dV maps in Figure 4a and 4b of the main manuscript, were performed around the energy position of these two resonances. 

\newpage 
\subsection{dI/dV point spectra and maps of pore states on trimer (4):} 
\begin{figure}[h] 
 	\vspace{1cm}
 \begin{center}
 	\includegraphics[width=0.95\textwidth]{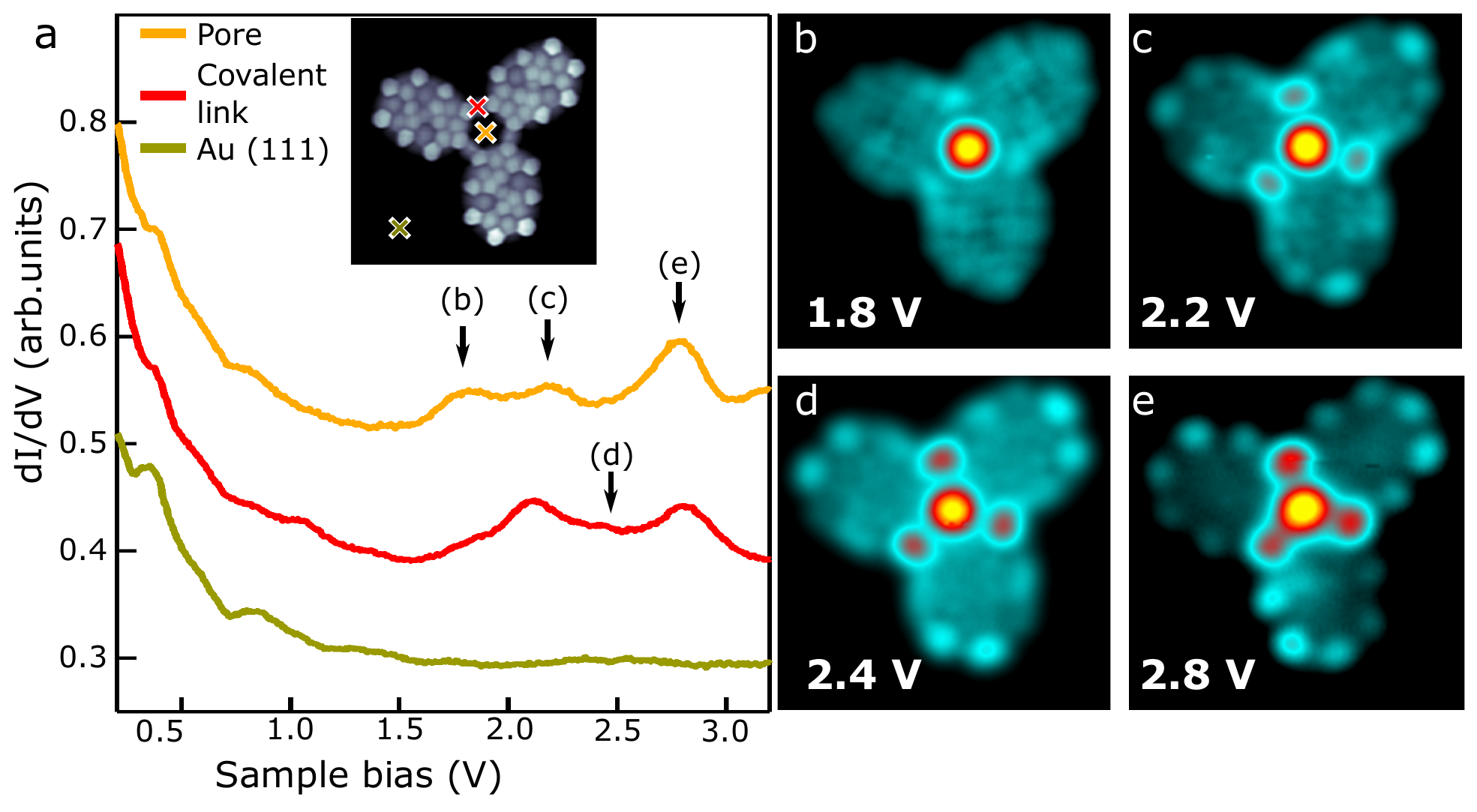}%\\
	\end{center}
 	\caption{STS constant current (\It=0.2~nA) point  spectra and corresponding   dI/dV maps showing the evolution of the pore state of a symmetric trimer \textbf{4}: (a) constant current dI/dV spectra taken at the pore (orange), at the covalent link (red). The spectrum of the Au(111) surface is shown as a reference (green); (b-e) constant height dI/dV maps taken with a metallic tip at the characteristic energies observed in a. \label{SN-STS-pore}} 	
 \end{figure}

To explore high-lying molecular states (i.e. at large sample bias) we use constant-current  dI/dV spectra (Fig- \ref{SN-STS-pore}a).  The spectra show several features at high bias, specially when acquired over the [18]annulene core of the molecule, which are attributed to molecular states localized at the pore cavity,  supported by constant-height dI/dV maps at the indicated bias values (Figure \ref{SN-STS-pore}(b-e)). 

\newpage
\subsection{dI/dV maps of dimer structures:} 
\begin{figure}[h]
 	 	\vspace{1cm}
 \begin{center}
 	\includegraphics[width=0.65\textwidth]{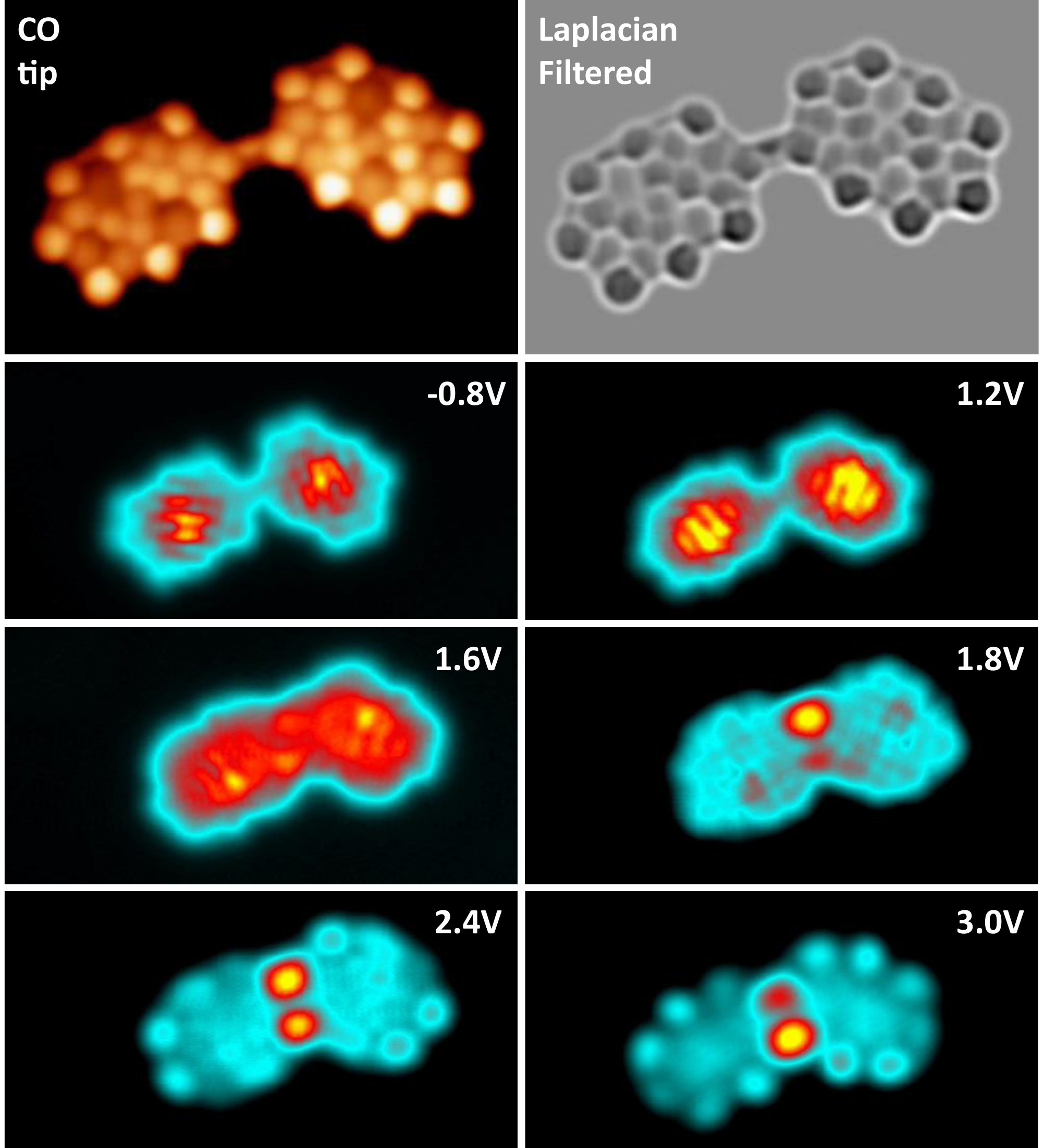}%\\
  \end{center}
 	\caption{STM image (\vs=5~mV, \It=1~nA) and corresponding dI/dV map of the dimer structure, taken with a CO functionalized tip. The conductance maps were taken in the constant height mode. %(images size 3.2$\times$3.9~nm$^2$, maps size 3.1$\times$2~nm$^2$). 
 	\label{SN-STS-dimer}}
 	\end{figure}

A small amount of dimers were also found over the sample. Fig. \ref{SN-STS-dimer} shows high resolution STM images  and constant height dI/dV spectral maps of one of such structures. We note that at high bias, precursor states of pore resonances appear localized at the semi-cavities at the connecting point. This constitutes a proof that the pore states are indeed molecular states instead of confined surface states.

\newpage

\addcontentsline{toc}{section}{Note 4: Density Functional Simulations of free trimer structures}
\section*{Note 4: Density Functional Simulations of free trimer structures} \label{SN-DFT}

To interpret the electronic properties of the trimers 
we performed first-principle density functional theory (DFT) simulations of the atomic structure
and electronic configuration of free-standing compound \textbf{4}.
High-resolution experimental observations allow us to conclude that the trimers adopt a planar conformation when adsorbed on the
Au(111) surface. For this reason,
despite not being the energetically most favorable conformation in gas phase,
 we have considered a planar trimer for our calculations, shown in  Figure \ref{DFT-EPS}a.

\subsection{Electrostatic Potential Energy Landscape:} \label{SN-DFT-EPS}

The experimental high resolution images show a clear contrast, with some of the peripheral rings appearing brighter 
and the 7-membered rings darker (Fig. 3a of the manuscript). It has been proposed that such intramolecular contrast could be related to an inhomogeneous electron density in the aromatic platform, which eventually could be associated to the stabilization of a Clar sextet structure in the graphenoid system \cite{Dienel2015}.
In turn, the inhomogeneous charge distribution affects the electrostatic potential landscape, which is known to have a significant influence on the features measured by high-resolution STM with a CO tip \cite{Hapala2016}. 
% Compound \textbf{4} is additionally distorted, and combines rings of different size.

\begin{figure}  [h]
	\begin{center}
		\includegraphics[width=0.95\textwidth]{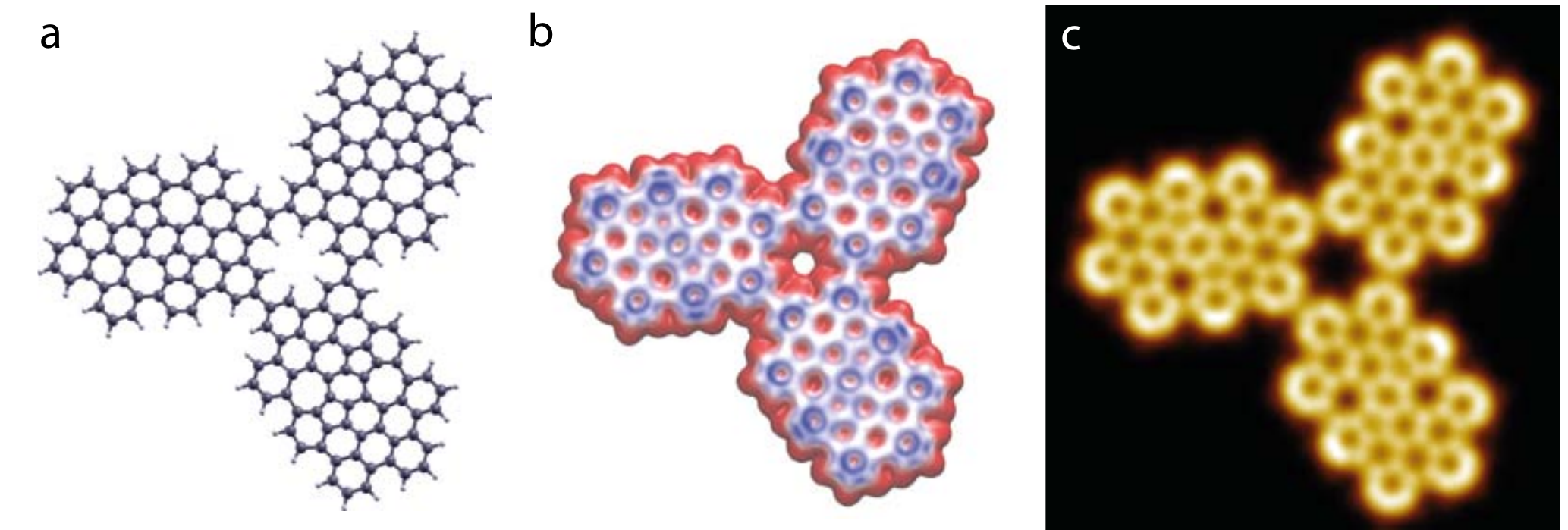}%\\
	\end{center}
	\caption{(a) Relaxed structure of free-standing compound 4, (b) charge density isosurface at 0.03 e/\AA$^3$ with a color range of [-272,272] meV, (c) constant height density of states at $\sim$3.0\AA,  above the molecular plane. 
		\label{DFT-EPS}}
\end{figure} 

In order to access to electrostatic potential energy landscapes,
we  simulated in Fig. \ref{DFT-EPS}b charge density isosurfaces   of the free trimer \textbf{4}. Clearly, a larger electron accumulation (more blue color) and, accordingly, a larger (more repulsive) electrostatic  potential is observed at the peripherial rings. In contrast, over  the 7-membered rings we find less charge density (more red color).   Such intramolecular differences  can also be detected   
in  constant height maps of the total density of states (Fig. \ref{DFT-EPS}c).
Thus, our DFT results  agree with the  interpretation of the "electronic" origin of the contrast in the high-resolution STM images of the trimer.

\subsection{Pore super-atom states:} \label{SN-SAMO}

%The electronic configuration of the free compound \textbf{4} was studied.

\begin{figure} [h]
	\begin{center}
		\includegraphics[width=0.55\textwidth]{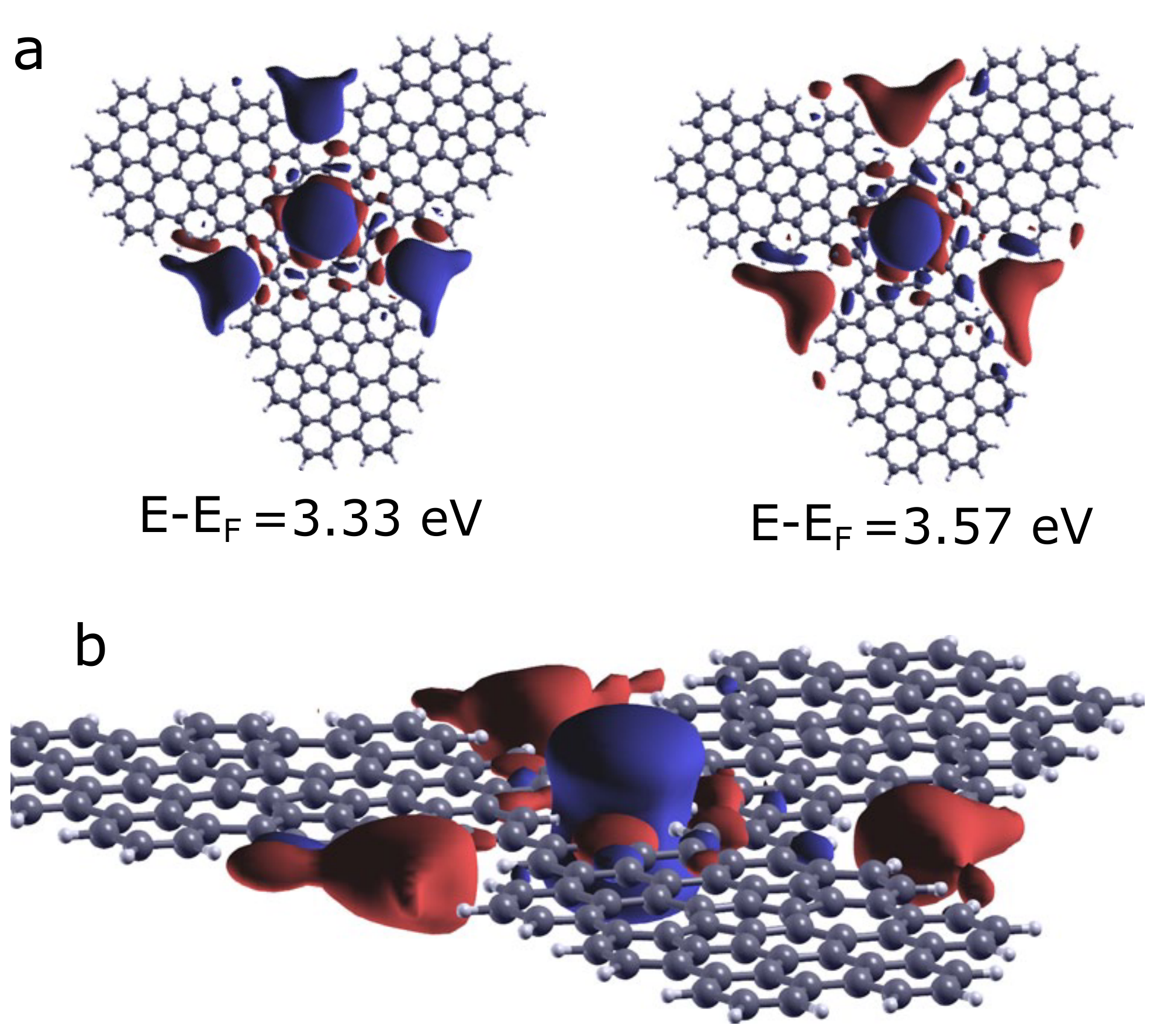}%\\
	\end{center}
	\caption{(a) Wavefunction amplitude isosurfaces (0.02\,electrons/bohr$^3$) of two states,
		with large weight at the [18]annulene pore. The color indicates the relative phase. (b) Wavefunction amplitude isosurface
		(0.03\,electrons/bohr$^3$)  of the state at 3.57 eV showing the even parity of this state with respect to the carbon plane.  \label{DFT-pore}}
\end{figure}

DFT finds at high energy above the Fermi level a set of states with larger weight in the [18]annulene  core and in the three semipores between the trimer blades. Figure \ref{DFT-pore}a shows two of these states appearing at 3.33\,eV (left) and 3.57\,eV above E$_F$ (right).  In both cases, the wave function is even with respect to the carbon plane (Figure \ref{DFT-pore}b). This proofs that the pore states found in the experiment are Super Atom Molecular Orbitals as defined by Hu and coworkers  \cite{Hu2010}. 

\newpage
\section*{References}
\vspace{1cm} 
 \bibliography{chem-bib}